*Article*

# Time-reversal Symmetry in Antenna Theory

**Mário G. Silveirinha**[1,*]

[1] University of Lisbon, Instituto Superior Técnico and Instituto de Telecomunicações, Avenida Rovisco Pais, 1, 1049-001 Lisboa, Portugal
\* Correspondence: mario.silveirinha@co.it.pt;

**Abstract:** Here, I discuss some implications of the time-reversal invariance of lossless radiating systems. I highlight that time-reversal symmetry provides a rather intuitive explanation for the conditions of polarization and impedance matching of a receiving antenna. Furthermore, I describe a solution to generate the time-reversed electromagnetic field through the illumination of a matched receiving antenna with a Herglotz wave.

**1. Introduction**

Time-reversal is the operation that flips the arrow of time such that $t \to -t$ [1, 2]. Remarkably, the laws that rule the microscopic dynamics of most physical systems are invariant under a time-reversal transformation (the exceptions occur in some nuclear interactions and are evidently irrelevant in context of this study). This property implies that under suitable initial conditions, the time reversed dynamics may be generated and observed in a real physical setting, similar to a movie played backwards. Ultimately, the invariance under time-reversal implies that at a microscopic level the physical phenomena are intrinsically reversible: if a particular time evolution is compatible with the physical laws, then the time-reversed dynamics also is.

A consequence of "time-reversal invariance" is that the propagation of light in standard waveguides is inherently bi-directional, even if the system does not have any particular spatial symmetry. For example, if some electromagnetic wave can go through some metallic pipe with no back-reflections, then the time-reversed wave also can, but propagating in the opposite direction. This rather remarkable property is usually explained with the help of the Lorentz reciprocity theorem [3-4], but it is ultimately a consequence of microscopic reversibility and time reversal invariance [2, 5].

In this article, I reexamine the consequences of time-reversal invariance in antenna theory. I show that time-reversal invariance provides a rather intuitive explanation for the conditions of impedance and polarization matching in the theory of the receiving antenna. In addition, I prove that in time-harmonic regime the time-reversed wave can be generated through the illumination of the receiving antenna with a superposition of plane waves generated in the far-field.

**2. Time-reversal symmetry**

It is well known that the equations of *macroscopic* electrodynamics are not time reversal invariant when the system has dissipative elements. This is so because the description provided by macroscopic electrodynamics is incomplete, as it only models the time evolution of the electromagnetic degrees of freedom. In contrast, in a microscopic formalism –with all the light and matter degrees of freedom included in the analysis– the system dynamics is time-reversal symmetric. Thus, in some sense, macroscopic dissipative systems (e.g., lossy dielectrics) have a *hidden* time-reversal symmetry. To circumvent this complication, here I will focus on systems with negligible material absorption, so that the dynamics determined by the macroscopic Maxwell equations is time-reversal symmetric.

*2.1. General case*

Consider the propagation of light in some lossless, dispersion-free, dielectric system described by the Maxwell equations:

$$\nabla \times \mathbf{E} = -\mu_0 \frac{\partial \mathbf{H}}{\partial t}, \qquad \nabla \times \mathbf{H} = \mathbf{j} + \varepsilon \frac{\partial \mathbf{E}}{\partial t}, \qquad (1)$$

with $\varepsilon = \varepsilon(\mathbf{r})$. The time-reversal operation $\mathcal{T}$ transforms the electromagnetic fields $\mathbf{E}, \mathbf{H}$ and the current density $\mathbf{j}$ as [2]:

$$\mathbf{E}(\mathbf{r},t) \xrightarrow{\mathcal{T}} \mathbf{E}(\mathbf{r},-t), \qquad \mathbf{H}(\mathbf{r},t) \xrightarrow{\mathcal{T}} -\mathbf{H}(\mathbf{r},-t), \qquad \mathbf{j}(\mathbf{r},t) \xrightarrow{\mathcal{T}} -\mathbf{j}(\mathbf{r},-t). \qquad (2)$$

The transformed fields satisfy the same equations as the original fields. Under a time-reversal transformation the magnetic field and the current density flip sign, whereas the electric field does not. Thus, the former are said to be odd under a time-reversal transformation, whereas the latter is even. As a consequence, the Poynting vector $\mathbf{S} = \mathbf{E} \times \mathbf{H}$ also flips sign under a time-reversal operation, as it should be so that the wave dynamics and direction of propagation are effectively reversed. The time reversal symmetry is rather general and applies to waves with an *arbitrary* variation in time.

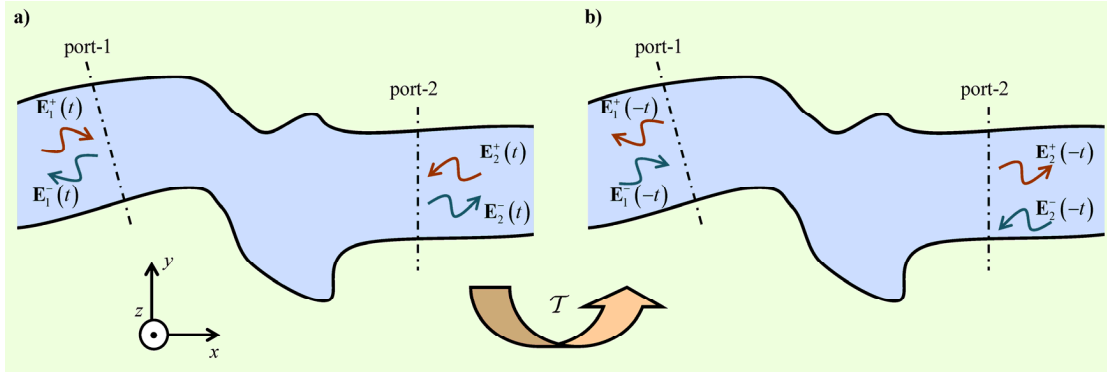

**Figure 1.** Illustration of the effect of the time-reversal operation in a time-domain scattering problem: (**a**) The incoming waves $\mathbf{E}_1^+$ and $\mathbf{E}_2^+$ are scattered by the junction and originate two outgoing waves $\mathbf{E}_1^-$ and $\mathbf{E}_2^-$. (**b**) Time-reversed scenario where the roles of the incoming and outgoing waves are exchanged.

For example, consider the scenario illustrated in Fig. 1a, which represents a scattering problem with two waveguides connected by some arbitrary junction (two-port microwave network). The two incoming waves $\mathbf{E}_1^+$ and $\mathbf{E}_2^+$ can have arbitrary time variations and their scattering originates two outgoing waves, $\mathbf{E}_1^-$ and $\mathbf{E}_2^-$. As illustrated in Fig. 1b, the time-reversal operation swaps the roles of the incoming and outgoing waves, because it flips the direction of propagation. Hence, the time-reversed signals are given by $\mathbf{E}_i^{\text{TR},\pm}(\mathbf{r},t) = \mathbf{E}_i^{\mp}(\mathbf{r},-t)$. In particular, suppose that some wave incident in port 1 is fully transmitted to port 2. Then, if port 2 is illuminated with the time-reversed transmitted signal it will reproduce the original signal in port 1, but reversed in time. Thereby, time-reversal invariant systems are intrinsically bi-directional, independent of any spatial asymmetry.

The enunciated results can be generalized in a straightforward way to dispersive lossless dielectrics, e.g., to material structures characterized by some real-valued scalar permittivity $\varepsilon = \varepsilon(\omega, \mathbf{r})$ (e.g., a Lorentz dispersive model with no dissipation). The reason is that the electrodynamics of lossless dispersive systems can be formulated as a Schrödinger-type time evolution problem [6-8], which in the case of reciprocal media (e.g., standard dielectrics) is time-reversal invariant.

Furthermore, as discussed in [9], most *lossless* nonlinear systems are time-reversal symmetric and hence are also bi-directional. For example, for an instantaneous Kerr-type nonlinear response with $\varepsilon = \varepsilon_0 \left( \chi^{(1)} + \chi^{(3)} \mathbf{E} \cdot \mathbf{E} \right)$, the Maxwell equations (1) remain time-reversal invariant. Interestingly, the time-harmonic response of a two-port microwave network with nonlinear components is generically asymmetric [10, 11, 12]. Indeed, if the ports are individually excited by the same time-harmonic signal the level of the transmitted signal depends on which port is excited; thus nonlinear systems are nonreciprocal [10, 11, 12]. In summary, lossless nonlinear systems are usually both time-reversal invariant and nonreciprocal, the two conditions are not incompatible [9]. In the rest of the article, I focus on linear systems.

*2.2. Time-harmonic variation*

Consider a time-harmonic solution of the Maxwell equations, such that the electromagnetic fields and current density are of the form: $\mathbf{E}(\mathbf{r},t) = \text{Re}\{\mathbf{E}_\omega(\mathbf{r}) e^{j\omega t}\}$, $\mathbf{H}(\mathbf{r},t) = \text{Re}\{\mathbf{H}_\omega(\mathbf{r}) e^{j\omega t}\}$, $\mathbf{j}(\mathbf{r},t) = \text{Re}\{\mathbf{j}_\omega(\mathbf{r}) e^{j\omega t}\}$, with $\omega$ the real-valued oscillation frequency. Under a time-reversal the electric field is transformed as $\mathbf{E}(\mathbf{r},t) \to \text{Re}\{\mathbf{E}_\omega(\mathbf{r}) e^{-j\omega t}\} = \text{Re}\{\mathbf{E}_\omega^*(\mathbf{r}) e^{j\omega t}\}$, where the symbol "*" stands for complex conjugation. Hence, the complex amplitudes of the fields and current density are transformed as:

$$\mathbf{E}_\omega(\mathbf{r}) \xrightarrow{\mathcal{T}} \mathbf{E}_\omega^*(\mathbf{r}), \qquad \mathbf{H}_\omega(\mathbf{r}) \xrightarrow{\mathcal{T}} -\mathbf{H}_\omega^*(\mathbf{r}), \qquad \mathbf{j}_\omega(\mathbf{r}) \xrightarrow{\mathcal{T}} -\mathbf{j}_\omega^*(\mathbf{r}). \tag{3}$$

Thus, in the frequency domain the time-reversal operation is closely linked to phase conjugation [13-14]. Similarly, voltages and currents are transformed as:

$$V_\omega \xrightarrow{\mathcal{T}} V_\omega^*, \qquad I_\omega \xrightarrow{\mathcal{T}} -I_\omega^*. \tag{4}$$

For example, consider a *N*-port microwave network such that the voltages and currents at a generic port *i* are of the form: $V_{\omega,i} = V_{\omega,i}^+ + V_{\omega,i}^-$ and $I_{\omega,i} = \left( V_{\omega,i}^+ - V_{\omega,i}^- \right) / Z_0$, $i=1,\ldots,N$. Here, $V_{\omega,i}^+$ represents an incoming (incident) wave and $V_{\omega,i}^-$ an outgoing (scattered) wave. The characteristic impedance of the ports is $Z_0$. The incident and scattered waves are related as $\mathbf{V}^- = \mathbf{S} \cdot \mathbf{V}^+$, where $\mathbf{V}^\pm = \left[ V_{\omega,i}^\pm \right]$ are column vectors and $\mathbf{S} = \left[ S_{ij} \right]$ is the scattering matrix. The time reversal operation exchanges the roles of the incident and scattered waves, such that $\mathbf{V}^{\text{TR},\pm} = \mathbf{V}^{\mp,*}$. Therefore, if the system is time-reversal invariant $\mathbf{V}^{+,*} = \mathbf{S} \cdot \mathbf{V}^{-,*}$. Thus, the scattering matrix must satisfy $\mathbf{S} = \mathbf{S}^{-1,*}$. On the other hand, for a lossless system the incident power must equal to the scattered power: $\mathbf{V}^- \cdot \mathbf{V}^{-,*} = \mathbf{V}^+ \cdot \mathbf{V}^{+,*}$. To satisfy this additional constraint the scattering matrix must be unitary $\mathbf{S} \cdot \mathbf{S}^\dagger = \mathbf{1}$. Combining the two results, one finds that the scattering matrix must be transpose symmetric:

$$\mathbf{S} = \mathbf{S}^T. \tag{5}$$

Thus, any time-reversal invariant linear lossless system is necessarily reciprocal ($S_{ij} = S_{ji}$) [15].

Here, I note in passing that in electromagnetic theory the time-reversal operator $\mathcal{T}$ is idempotent, such that $\mathcal{T}^2 = \mathbf{1}$. In contrast, in condensed matter theory the time-reversal operator satisfies $\mathcal{T}^2 = -\mathbf{1}$, and because of this property the scattering matrix of fermionic systems is anti-symmetric, $\mathbf{S} = -\mathbf{S}^T$ [15]. It was recently shown that photonic systems protected by a special parity-time-duality ($\mathcal{PTD}$) symmetry are constrained by $\mathbf{S} = -\mathbf{S}^T$, and thereby are matched at all ports ($S_{ii} = 0$) [15]. Such systems can enable bi-directional transmission of light free of back-scattering.

**3. Application to antenna theory**

The time-reversal property may be used to explain several well known properties of radiating systems. Similar to the previous section, I assume that the antennas are formed by lossless materials, e.g., lossless dielectrics or perfect conductors. In particular, the radiation efficiency of the antennas is 100%.

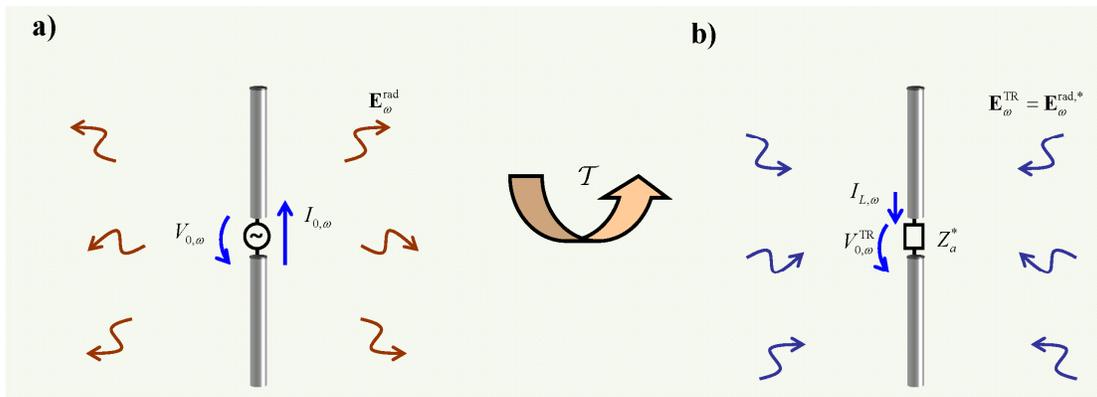

**Figure 2.** a) An antenna fed by a time-harmonic generator radiates in free-space. b) Time-reversed problem wherein all the radiated energy returns to the antenna. The antenna terminals are connected to a matched load.

Consider a generic antenna radiating in free-space (Fig. 2a). The antenna is fed by a generator with a time-harmonic variation. The antenna radiates the electromagnetic fields $\mathbf{E}_\omega^{\rm rad}, \mathbf{H}_\omega^{\rm rad}$. By definition, the antenna impedance is $Z_a = V_{0,\omega}/I_{0,\omega}$ where $V_{0,\omega}, I_{0,\omega}$ are the complex amplitudes of the voltage and current at the antenna terminals. In the far-field region the radiated electric field is asymptotically of the form [16]:

$$\mathbf{E}_\omega^{\rm rad} \approx \mathbf{E}_\omega^{\rm ff} \equiv \eta_0 j k_0 I_{0,\omega} \frac{e^{-jk_0 r}}{4\pi r} \mathbf{h}_e(\hat{\mathbf{r}}), \quad \text{with} \quad \mathbf{h}_e(\hat{\mathbf{r}}) = \hat{\mathbf{r}} \times \left( \hat{\mathbf{r}} \times \frac{1}{I_{0,\omega}} \int \mathbf{j}_\omega(\mathbf{r}') e^{jk_0 \hat{\mathbf{r}}\cdot\mathbf{r}'} d^3\mathbf{r}' \right). \tag{6}$$

In the above, $k_0 = \omega/c$ is the free-space wave number, $\eta_0$ is the free-space impedance, and $\mathbf{h}_e(\hat{\mathbf{r}})$ is the (vector) effective height of the antenna, which depends on the direction of observation $\hat{\mathbf{r}}$ ($\hat{\mathbf{r}}$ can be expressed in terms of angles $\theta, \varphi$ associated with a system of spherical coordinates). The antenna effective height depends on the total current distribution $\mathbf{j}_\omega(\mathbf{r}')$, which includes the external currents associated with the generator and the induced polarization and conduction currents in the materials.

The polarization of the antenna in the direction $\hat{\mathbf{r}}$ is determined by the closed curve defined by $\mathbf{E}^{\rm rad}(t) = {\rm Re}\{\mathbf{E}_\omega^{\rm rad} e^{j\omega t}\} \sim {\rm Re}\{\mathbf{h}_e(\hat{\mathbf{r}}) e^{j\omega t}\}$, and hence by the effective height $\mathbf{h}_e$ because the electric field is evaluated in the far-field region.

*3.1. Polarization and impedance matching*

Consider now the time-reversed problem represented in Fig. 2b, where all the radiated energy is returned back to the antenna. The time reversed voltage and current at the antenna terminals are $V_{0,\omega}^{\rm TR} = V_{0,\omega}^*$ and $I_{0,\omega}^{\rm TR} = -I_{0,\omega}^*$ [Eq. (4)]. The current flowing into the antenna terminals (inward direction) is $I_{L,\omega} = -I_{0,\omega}^{\rm TR}$ (see Fig. 2; note that in the scenario of Fig. 2a the current is positive when it flows in the outward direction). From here, it follows that $V_{0,\omega}^{\rm TR}/I_{L,\omega} = V_{0,\omega}^*/I_{0,\omega}^* = Z_a^*$, i.e., in the time-reversed scenario the generator is effectively equivalent to a *matched load* with impedance $Z_a^*$. Furthermore, the field arriving to the antenna in the direction $\hat{\mathbf{r}}$ is evidently $\mathbf{E}_\omega^{\rm TR} = \mathbf{E}_\omega^{\rm rad,*} \sim \mathbf{h}_e^*(\hat{\mathbf{r}})$, which is the well-know condition for *polarization matching*. These properties show that in the time-reversed problem the antenna is *impedance matched* to the load and *polarization matched* to the incident wave for any direction $\hat{\mathbf{r}}$.

Thus, the time-reversal invariance provides a rather intuitive understanding of the conditions of impedance and polarization matching, as it shows that the two conditions emerge naturally in the time-reversed problem where the receiving antenna captures with 100% efficiency the energy arriving from the far-field.

In the time domain $\mathbf{E}^{TR}(t) = \mathbf{E}^{rad}(-t)$, and thereby the polarization curve associated to $\mathbf{E}_\omega^{TR}$ is the same as the polarization curve associated with $\mathbf{E}_\omega^{rad}$. In a time period ($T = 2\pi/\omega$), $\mathbf{E}^{TR}(t), \mathbf{E}^{rad}(t)$ follow the same polarization curve but in *opposite* directions due to the time-reversal link. Yet, the polarization of the two waves is the same, i.e., the antenna and the wave are polarization matched, because the propagation directions of the two waves differ by a minus sign ($\mathbf{E}_\omega^{rad}$ propagates in the outward radial direction and $\mathbf{E}_\omega^{TR}$ in the inward radial direction).

For example, an antenna that radiates a right-circularly polarized (RCP) wave in some direction of space is polarization matched to an incoming plane wave with RCP polarization. Even though the wave and antenna polarizations are identical, the geometrical senses of rotation of the relevant electric fields are *opposite*. This otherwise intriguing property can be understood as a simple consequence of time-reversal invariance.

*3.2. Time-reversed field generated with a far-field illumination*

The problem of generating a time-reversed field distribution is of practical interest, as it enables concentrating and focusing energy from the far-field into some desired region of space. The theory and application of time-reversed fields were developed and extensively explored by Fink and co-authors [17-21]. Here, I revisit the problem and highlight some features that were not discussed in Ref. [19].

In the time-reversed problem of Fig. 2b the incident wave $\mathbf{E}_\omega^{TR}$ propagates from $r = \infty$ to the antenna where it is fully absorbed by the matched load, without generating any back-reflections. It is natural to wonder what happens if the same antenna is illuminated by the time-reversed far-field (time reversal of $\mathbf{E}_\omega^{ff}$) rather than by the fully time-reversed field (the time reversal of $\mathbf{E}_\omega^{rad}$ given by $\mathbf{E}_\omega^{TR}$). In the former case, the incident wave $\mathbf{E}_\omega^{inc}(\mathbf{r})$ should be a superposition of plane waves emerging from all possible directions of space $\hat{\mathbf{r}}'$. From Eq. (6), the field $d\mathbf{E}_\omega^{inc}(\mathbf{r})$ associated with the wave emerging from the infinitesimal solid angle $d\Omega(\hat{\mathbf{r}}')$ must have amplitude proportional to $e^{+jk_0\hat{\mathbf{r}}'\cdot\mathbf{r}}\mathbf{h}_e^*(\hat{\mathbf{r}}')d\Omega(\hat{\mathbf{r}}')$.

Notably, I prove in the Appendix that the solution of the scattering problem formulated in the previous paragraph can be constructed from the fully-time reversed field $\mathbf{E}_\omega^{TR}$. Specifically, when an *impedance-matched* antenna is illuminated by the incident field

$$\mathbf{E}_\omega^{inc}(\mathbf{r}) = \frac{k_0^2}{8\pi^2} I_{0,\omega}^* \eta_0 \int e^{+jk_0\hat{\mathbf{r}}'\cdot\mathbf{r}} \mathbf{h}_e^*(\hat{\mathbf{r}}') d\Omega(\hat{\mathbf{r}}'), \qquad (7)$$

the field scattered by the antenna is precisely given by $\mathbf{E}_\omega^{scat} = \mathbf{E}_\omega^{TR} - \mathbf{E}_\omega^{inc}$, such that the total field is $\mathbf{E}_\omega^{TR}$. Thus, $\mathbf{E}_\omega^{TR}$ may be both understood as an incident wave that is absorbed by the antenna with no back-scattering, or alternatively as the superposition of an incident wave ($\mathbf{E}_\omega^{inc}$) and the corresponding field back-scattered by the antenna ($\mathbf{E}_\omega^{scat}$). The two cases, even though totally different from a physical point of view, cannot be mathematically distinguished in time-harmonic regime.

As previously mentioned, the incident field $\mathbf{E}_\omega^{inc}$ is a superposition of propagating plane waves emerging from all directions of space. This type of wave is known as a Herglotz wave. The integral in Eq. (7) is over all solid angles $d\Omega(\hat{\mathbf{r}}')$. Furthermore, it is shown in Appendix A that the scattered field has the following asymptotic form in the far-field region:

$$\mathbf{E}_\omega^{\text{scat}}(\mathbf{r}) \approx -\eta_0 jk_0 I_{0,\omega}^* \frac{e^{-jk_0 r}}{4\pi r} \mathbf{h}_e^*(-\hat{\mathbf{r}}). \tag{8}$$

Comparing Eqs. (6) and (8) it is evident that the power scattered by the antenna when it is illuminated by $\mathbf{E}_\omega^{\text{inc}}$ is $P_{\text{scat}} = P_{\text{rad}}$ where $P_{\text{rad}}$ is the power radiated by the antenna in the scenario of Fig. 2a. Furthermore, since the total field $\mathbf{E}_\omega^{\text{inc}} + \mathbf{E}_\omega^{\text{scat}}$ is identical to $\mathbf{E}_\omega^{\text{TR}}$ it is evident that the power absorbed by the matched load is $P_r = P_{\text{rad}}$. These properties imply that when the impedance matched antenna is illuminated by the Herglotz wave it captures the same power as it scatters: $P_r = P_{\text{scat}}$. The property $P_r = P_{\text{scat}}$ is specific of the Herglotz wave considered here, and it generally does not hold true for other far-field excitations [22-23]. Note that the polarization curve associated with the scattered field $\mathbf{E}_\omega^{\text{scat}}$ along the direction $\hat{\mathbf{r}}$ is determined by $\mathbf{h}_e^*(-\hat{\mathbf{r}})$, which generally differs from the polarization of the antenna in transmitting mode.

It is emphasized that the fully-time reversed field $\mathbf{E}_\omega^{\text{TR}}$ can be excited simply by illuminating the impedance matched antenna with the Herglotz wave $\mathbf{E}_\omega^{\text{inc}}$, which is a superposition of propagating plane waves.

## 4. Conclusions

I revisited the topic of time-reversal symmetry in macroscopic electromagnetism. I showed that under a time-reversal transformation a transmitting antenna becomes the impedance matched receiving antenna. Heuristically, the excitation with the time-reversed wave must be the most effective way of delivering power to an antenna. Thus, the time-reversal invariance provides a simple and intuitive understanding of the conditions of impedance and polarization matching in antenna theory. In particular, it elucidates why a polarization matched incident wave has an electric field that rotates geometrically in a direction opposite to that of the field radiated by the antenna in the same direction. In addition, I generalized the ideas of [19] and showed that the time-reversal of the field emitted by a lossless transmitting antenna can be created by illuminating an impedance matched receiving antenna with the far-field excitation associated with the Herglotz wave (7). In such a scenario, the power captured by the matched load is precisely the same as the power scattered by the antenna.

**Funding:** This research was funded by the IET under the A F Harvey Engineering Research Prize and by Fundação para Ciência e a Tecnologia (FCT) under project PTDC/EEITEL/ 4543/2014 and UID/EEA/50008/2017.

**Conflicts of Interest:** The authors declare no conflict of interest.

## Appendix A

Consider the configuration of Fig. 2a, where a generic lossless antenna radiates in free-space. Let $\mathbf{j}_\omega(\mathbf{r}')$ be the total electric current distribution, determined both by the external current associated with generator and by the polarization and conduction currents in the materials. The radiated fields in time-harmonic regime may be expressed in terms of a vector potential as

$$\mathbf{A}_\omega(\mathbf{r}) = \mu_0 \int \mathbf{j}_\omega(\mathbf{r}') \frac{e^{-jk_0|\mathbf{r}-\mathbf{r}'|}}{4\pi|\mathbf{r}-\mathbf{r}'|} d^3\mathbf{r}'. \tag{A1}$$

Under a time-reversal, the vector potential is transformed as $\mathbf{A}_\omega(\mathbf{r}) \xrightarrow{T} -\mathbf{A}_\omega^*(\mathbf{r})$. Thus, the time-reversed vector potential is:

$$\mathbf{A}_\omega^{\text{TR}}(\mathbf{r}) = \mu_0 \int -\mathbf{j}_\omega^*(\mathbf{r}') \frac{e^{+jk_0|\mathbf{r}-\mathbf{r}'|}}{4\pi|\mathbf{r}-\mathbf{r}'|} d^3\mathbf{r}'. \tag{A2}$$

Using $e^{+jk_0|\mathbf{r}-\mathbf{r}'|} = e^{-jk_0|\mathbf{r}-\mathbf{r}'|} + 2j\sin(k_0|\mathbf{r}-\mathbf{r}'|)$, I obtain the decomposition $\mathbf{A}_\omega^{\text{TR}} = \mathbf{A}_\omega^{\text{inc}} + \mathbf{A}_\omega^{\text{scat}}$, with

$$\mathbf{A}_\omega^{scat}(\mathbf{r}) = \mu_0 \int -\mathbf{j}_\omega^*(\mathbf{r}') \frac{e^{-jk_0|\mathbf{r}-\mathbf{r}'|}}{4\pi|\mathbf{r}-\mathbf{r}'|} d^3\mathbf{r}', \tag{A3a}$$

$$\mathbf{A}_\omega^{inc}(\mathbf{r}) = \mu_0 \int -\mathbf{j}_\omega^*(\mathbf{r}') \frac{j\sin(k_0|\mathbf{r}-\mathbf{r}'|)}{2\pi|\mathbf{r}-\mathbf{r}'|} d^3\mathbf{r}'. \tag{A3b}$$

Evidently, the time-reversed field has a similar decomposition $\mathbf{E}_\omega^{TR} = \mathbf{E}_\omega^{inc} + \mathbf{E}_\omega^{scat}$ (see also Ref. [19]). The field $\mathbf{E}_\omega^{scat}$ is obtained from $\mathbf{A}_\omega^{scat}$, and thus satisfies the Sommerfeld radiation conditions. Thus, $\mathbf{E}_\omega^{scat}$ can be understood as the wave scattered by $\mathbf{E}_\omega^{inc}$. From Eq. (A3a) it is simple to check that in the far-field region

$$\mathbf{E}_\omega^{scat}(\mathbf{r}) \approx -jk_0\eta_0 \frac{e^{-jk_0 r}}{4\pi r} \hat{\mathbf{r}} \times \left( \hat{\mathbf{r}} \times \int \mathbf{j}_\omega^*(\mathbf{r}') e^{+jk_0\hat{\mathbf{r}}\cdot\mathbf{r}'} d^3\mathbf{r}' \right). \tag{A4}$$

Comparing this result with Eq. (6), one obtains Eq. (8).

The potential $\mathbf{A}_\omega^{inc}$ is an analytic function and can be written as a superposition of plane waves. Indeed, from

$$\frac{\sin k_0 r}{4\pi r} = \frac{k_0}{16\pi^2} \int e^{-jk_0\hat{\mathbf{k}}\cdot\mathbf{r}} d\Omega(\hat{\mathbf{k}}), \tag{A5}$$

the incident vector potential may be expressed as:

$$\mathbf{A}_\omega^{inc}(\mathbf{r}) = -\mu_0 \frac{jk_0}{8\pi^2} \int d\Omega(\hat{\mathbf{k}}) e^{jk_0\hat{\mathbf{k}}\cdot\mathbf{r}} \left( \int d^3\mathbf{r}' \mathbf{j}_\omega^*(\mathbf{r}') e^{-jk_0\hat{\mathbf{k}}\cdot\mathbf{r}'} \right). \tag{A6}$$

With the help of Eq. (6), it can be checked that the "incident" electric field $\mathbf{E}_\omega^{inc} = (1/j\omega\varepsilon_0)\nabla\times\nabla\times\mathbf{A}_\omega^{inc}/\mu_0$ is given by Eq. (7).

The power received by an impedance matched antenna is

$$P_r = \frac{|V_{oc}|^2}{8R_a}. \tag{A7}$$

Here, $R_a = \text{Re}\{Z_a\}$ is the input resistance of the antenna and $V_{oc}$ is the voltage induced by the incident field at the antenna terminals when they are terminated with an open-circuit. As is well known, for reciprocal systems the open-circuit voltage is $V_{oc} = \mathbf{E}_0^{inc} \cdot \mathbf{h}_e(\hat{\mathbf{r}})$ where $\mathbf{E}_0^{inc}$ is field associated with an incident plane wave (arriving from direction $\hat{\mathbf{r}}$) evaluated at the origin [16]. Thus, from the superposition principle, the voltage induced by the Herglotz wave (7) is:

$$V_{oc} = \frac{k_0^2}{8\pi^2} I_{0,\omega}^* \eta_0 \int |\mathbf{h}_e(\hat{\mathbf{r}})|^2 d\Omega(\hat{\mathbf{r}}). \tag{A8}$$

For a lossless system the input resistance is coincident with the radiation resistance, which from (6) can be written as $R_a = \eta_0 \frac{k_0^2}{16\pi^2} \int |\mathbf{h}_e(\hat{\mathbf{r}})|^2 d\Omega(\hat{\mathbf{r}})$. This result implies that $V_{oc} = 2I_{0,\omega}^* R_a$ so that the received power is given by $P_r = \frac{1}{2} R_a |I_{0,\omega}|^2 = P_{rad}$. This direct analysis confirms that when the antenna is illuminated by the Herglotz wave the power absorbed by a matched load is exactly the power radiated by the antenna in the scenario of Fig. 2a.

**References**

1. Feynman, R.; Leighton, R.; Sand, M.; *The Feynman lectures on physics*, California Institute of Technology, 1963.
2. Casimir, H. B. G.; Reciprocity theorems and irreversible processes, *Proc. IEEE*, **1963**, 51, 1570.
3. Potton, R. J.; Reciprocity in optics, *Rep. Prog. Phys.*, **2004**, 67, 717-754.

4. Caloz, C.; Alù, A.; Tretyakov, S.; Sounas, D.; Achouri, K.; Deck-Léger, Z.-L.; Electromagnetic Nonreciprocity, *Phys. Rev. Applied*, **2018**, 10, 047001.
5. Onsager, L.; Reciprocal relations in irreversible processes I, **1931**, *Phys. Rev.* 37, 405.
6. Silveirinha, M. G.; Chern Invariants for Continuous Media, *Phys. Rev. B*, **2015**, 92, 125153.
7. Silveirinha, M. G.; Topological classification of Chern-type insulators by means of the photonic Green function", *Phys. Rev. B*, **2018**, 97, 115146.
8. Silveirinha, M. G.; "Modal expansions in dispersive material systems with application to quantum optics and topological photonics", chapter to appear in "Advances in Mathematical Methods for Electromagnetics", (edited by Paul Smith, Kazuya Kobayashi) IET, (available in arXiv:1712.04272).
9. Fernandes, D. E.; Silveirinha, M. G., Asymmetric Transmission and Isolation in Nonlinear Devices: Why They Are Different, *IEEE Antennas Wirel. Propag. Lett.*, **2018**, *17*, 1953.
10. Shadrivov, I. V.; Fedotov, V. A.; Powell, D. A.; Kivshar, Y. S.; Zheludev, N. I.; Electromagnetic wave analogue of an electronic diode, *New J. Phys.*, **2011**, 13, 033025.
11. Mahmoud, A. M., Davoyan, A. R.; Engheta, N.; All-passive nonreciprocal metastructure, *Nat. Commun.*, **2015**, 6, 8359.
12. Sounas, D. L.; Soric, J.; Alù, A.; Broadband Passive Isolators Based on Coupled Nonlinear Resonances, *Nat. Electron.*, **2018**, 1, 113-119.
13. Maslovski, S.; Tretyakov S.; Phase conjugation and perfect lensing, *J. Appl. Phys*, **2003**, *94*, 4241.
14. Pendry, J. B.; Time Reversal and Negative Refraction, *Science*, **2008**, *322*, 71-73.
15. Silveirinha, M. G.; PTD symmetry protected scattering anomaly in optics, *Phys. Rev. B*, **2017**, *95*, 035153.
16. Balanis, C. A.; *Antenna theory,* (4th Ed.) John Wiley & Sons, Inc., Hoboken, NJ, USA, 2016.
17. Lerosey, G.; Rosny, J. de; Tourin, A.; Derode, A.; Montaldo, G.; Fink, M.; Time Reversal of Electromagnetic Waves, *Phys. Rev. Lett.*, **2004**, 92, 193904.
18. Lerosey, G.; Rosny, J. de; Tourin, A.; Fink, M.; Focusing Beyond the Diffraction Limit with Far-Field Time Reversal, *Science*, **2007**, 315, 1120.
19. Rosny, J. de; Lerosey, G.; Fink, M.; Theory of Electromagnetic Time-Reversal Mirrors, *IEEE Trans. Antennas Propag.*, **2010**, 58, 3139-3149.
20. Fink, M.; Time-Reversed Acoustics, *Scientific American*, **1999**, 281, 91-97.
21. Rosny J. de; Fink, M.; Overcoming the Diffraction Limit in Wave Physics Using a Time-Reversal Mirror and a Novel Acoustic Sink, *Phys. Rev. Lett.*, **2002**, 89, 124301.
22. Andersen J. B.; Vaughan, R. G.; Transmitting, receiving and scattering properties of antennas, *IEEE Antennas Propag. Mag.*, **2003**, 45, 93–98.
23. Andersen, J. B.; Frandsen, A.; Absorption Efficiency of Receiving Antennas, *IEEE Trans. Antennas Propag.*, **2005**, 53, 2843-2849.